\documentstyle[12pt]{article}
\newcommand{\rlnabla}{\nabla^{{}^{{}^{\!\!\!\!\!\!\!\!\!
\longleftrightarrow}}}}
\newcommand{\be}{\begin{equation}}
\newcommand{\ee}{\end{equation}}
\begin{document}
\def\theequation{\arabic{section}.\arabic{equation}}
\begin{titlepage}
\title{Tales of tails in cosmology}
\author{Valerio Faraoni$^1$ and Edgard Gunzig$^{1,2}$\\ \\
{\small \it $^1$ RggR, Facult\`e des Sciences}\\
{\small \it Campus Plaine, Universit\`e Libre de Bruxelles}\\
{\small \it Boulevard du Triomphe, CP231}\\ 
{\small \it 1050 Bruxelles, Belgium}\\\\
{\small \it $^2$ Instituts Internationaux de Chimie et de Physique Solvay}
}
\date{}
\maketitle   
\thispagestyle{empty}  
\vspace*{1truecm}
\begin{abstract} 
Late time mild inflation (LTMI) proposes to solve the age of
the universe problem and the discrepancy between locally and
globally  measured values of the Hubble parameter. However, the mechanism
proposed to achieve LTMI is found to be physically pathological by
applying the theory of tails for the solutions of wave equations in curved
spaces. Alternative mechanisms for LTMI are discussed, and the relevance
of scalar wave tails for cosmology is emphasized.
\end{abstract}
\vspace*{1truecm} 
\begin{center} 
To appear in {\em Int. J. Mod. Phys. D}. 
\end{center}     
\end{titlepage}   \clearpage

\section{Introduction}

\setcounter{equation}{0}

The late time mild inflationary (LTMI) scenario has recently been proposed
\cite{LTMI} to solve the age of the universe problem, and the puzzle of
the discrepancy between the locally and globally measured values of the
Hubble parameter. 

In this paper, we discuss the homogeneous Klein--Gordon equation for
a scalar field from the point of view of the Huygens' principle and wave
tails, with respect to its cosmological applications. In the LTMI
scenario, the scalar field is allowed to couple explicitly to the Ricci
curvature of spacetime  (see Eq. (\ref{2}) below). The physical reasons to
consider a nonminimal coupling term are many, and are summarized in Ref.
\cite{F96PRD}; indeed, the nonminimal coupling is forced upon us by
the physics of the scalar field \cite{F96PRD}. The study of the Huygens'
principle and of scalar tails leads to unexpected physics
\cite{SF93CQG,GribPoberii,GribRodrigues,Waterloo}. The main goal of the
present paper is the application of these results to cosmology, in
particular to LTMI,
continuing the program initiated in Refs. 
\cite{FS92PLA,F96PRD,F97GRG,Calgary}.

An interesting issue in the physics of wave propagation is the validity of
the Huygens' principle. A field satisfying a linear wave equation can
propagate ``sharply'' along the characteristic surfaces, or with ``tails''
of
radiation, reverberations that degrade the information carried by a
initially delta--like pulse, and that violate the Huygens' principle. To
be clear,
we adopt the physical definition of Huygens' principle due to Hadamard 
\cite{Hadamard}. Assume that a delta--like pulse of radiation (light,
for example) is emitted
by a point--like source in $ P$, at the time $ t=0 $. If, at
the time $ t>0$, the radiation is entirely confined to the surface of the
sphere of center $P$
and radius $r=ct$ (where $c$ is the speed of light), one says that the
Huygens' principle is satisfied. If, on the contrary, there is radiation
at radii $r$ such that $ r<ct $, there are {\em tails} of radiation: the
waves
are spread at any radius. A precise mathematical 
definition\footnote{Unfortunately, the terminology commonly used in the 
literature is misleading; it would be more appropriate to refer to the
``Huygens' property'' instead of the ``Huygens' principle''.} is found in
Sec. 2.

It is known that, given a wave equation in a curved spacetime 
$ (M, g_{ab} ) $, the Huygens' principle is generally violated by
its solutions, due to the following possibilities: the presence of a mass
term in the wave equation satisfied by the field, the dimensionality of
spacetime, and 
backscattering off the background curvature of spacetime 
\cite{Hadamard}--\cite{Gunther}. The first of these causes is trivial and
well known. Moreover, in this paper there are no tails due to the
spacetime dimensionality. Backscattering off the spacetime curvature, on
the other hand, is nontrivial and the presence or absence of tails
for scalar, electromagnetic, and gravitational waves has been established
only for a handful of spacetime metrics $ g_{ab}  $.

It is not surprising that the study of violations of the Huygens'
principle has fruitful applications to cosmology, in view of the 
fact that scalar fields are widely used in this area, expecially in
inflationary theories of the early universe, and as candidates for dark
matter in today's universe. Also, it is worth reminding the reader that
tails of gravitational waves
due to the spacetime curvature near compact sources have received
attention in conjunction with the data analysis of the large
interferometric detectors of gravitational waves \cite{gwtails}. The
relevance of tails for cosmological gravitational waves is, instead,
unclear.

The plan of the paper is as follows: in Sec. 2, a theorem valid for
massive fields of spin $ s \geq 1/2 $
satisfying wave equations is recalled, and analogous results are derived
for the massive spin 0 field. The importance of a correct formulation of
the Huygens' principle for physical applications is emphasized. Then, we
proceed to study an ``ultrapathological'' case of wave propagation for a
scalar field in a curved space. In Sec. 3, which is the most relevant 
to cosmology,  the late time mild inflationary
scenario of the universe is studied, and it is shown
that this scenario essentially coincides with the ultrapathological space
of Sec. 2. Alternative mechanisms to achieve late time mild inflation are
discussed. Section 4 presents the conclusions. 

\section{Massive fields in curved spaces and the tail--free property}

\setcounter{equation}{0}
Massive fields of arbitrary spin satisfying wave equations in a curved
space have been studied for a long time, both from the mathematical and
the physical (classical and quantum) point of view. In this paper, we
restrict ourselves to the classical aspects of the physics of wave
propagation, in particular the violation of the Huygens' principle and the
occurrence of tails of radiation for a field satisfying a wave equation
\cite{Hadamard}--\cite{Gunther}.

It is required that the fields considered
live in the spacetime $(M,g_{ab})$, where $M$ is a four--dimensional
smooth manifold, $g_{ab}$ is the metric tensor, and $\nabla_{a} $ is
the associated covariant derivative operator\footnote{The metric signature
is --~+~+~+.  The speed of light and Planck's constant assume the value
unity.  The Ricci tensor is given by $R_{\mu\rho}= \Gamma^{\nu}_{\mu\rho
,\nu}-\Gamma^{\nu}_{\nu\rho ,\mu}+
\Gamma^{\alpha}_{\mu\rho}\Gamma^{\nu}_{\alpha\nu}-
\Gamma^{\alpha}_{\nu\rho}\Gamma^{\nu}_{\alpha\mu} $ in terms of the
Christoffel symbols $\Gamma_{\alpha\beta}^{\delta}$, and
$R={R_{\mu}}^{\mu}$. The abstract index notation is used.}. 

We begin by considering massive fields with spin $s \geq 1/2$, which have
recently been the subject of renewed interest \cite{Ilge}; the following
theorem is
valid (we refer the reader to Ref. \cite{Wunsch} for the relevant
equations and a proof):\\\\ {\bf Theorem 1:} {\em A solution of the
homogeneous wave equation for a massive field with spin $s\geq 1/2$ on the
spacetime $(M,g_{ab})$ obeys the Huygens' principle if and only if
$(M,g_{ab})$ is a spacetime of constant curvature and the Ricci scalar
satisfies \be \label{1} R=\frac{6m^2}{s} \; , \ee where $m$ is the mass of
the field.}\\\\ The formulation of the Huygens' principle used in Theorem
1 and in its proof is crucial. In fact, although the Huygens' principle
for the solutions of a wave equation was formulated by Hadamard
\cite{Hadamard} in a clear and physically meaningful way as the absence of
tails of radiation, several other
definitions have been introduced in the literature over the years: 
characteristic propagation property, progressing--wave propagation, etc. 
These definitions are {\em a priori} inequivalent, and they are all
loosely referred to as the ``Huygens' principle''. This improper
terminology is often a source of confusion and misinterpretations of
mathematical results (see Refs. \cite{SF92JMP,BombelliSonego} for a
clarification of the relationships between at least some of the various
definitions proposed in the literature). In the following, we consider the
analogue of Theorem 1 for the case of the massive scalar field ($s=0$). To
this end, we first provide a unambigous definition of the Huygens'
principle.

A scalar field $\phi$ in a source--free region of spacetime satisfies the
homogeneous Klein--Gordon equation 
\be \label{2} 
g^{ab}\nabla_a \nabla_b
\phi -m^2 \phi -\xi R \, \Phi =0\;, 
\ee 
where the dimensionless constant $
\xi $ describes the direct coupling between the field $\phi$ and the Ricci
curvature $R$ of spacetime. The formal solution of Eq.~(\ref{2}) is given
by a Green function representation in a normal domain $ \cal{N} $ of
spacetime not containing sources as \be \label{3} \phi (x) 
=\int\limits_{\partial \cal{N} }dS^{a'}(x') \, G\left( x',x \right)
{\rlnabla}_{a'} \, \phi (x') \;, \ee where $\partial \cal{N} $ is the
boundary of the normal domain $\cal{N}$, $ dS^{a'}(x') $ is the oriented
volume element on the hypersurface $\partial \cal{N} $ at $ x' $, and \be
\label{4} f_{1} \rlnabla f_{2} \equiv f_{1} \nabla f_{2}-f_{2}\nabla f_{1}
\ee for any pair of differentiable functions $ \left( f_{1} , f_{2}\right) 
$. For
physical reasons, we restrict ourselves to the consideration of the
retarded Green function $G_R \left( x',x \right) $, which is a solution of
the wave equation~(\ref{2}) with an impulsive source located at $x$, \be
\label{5} \left[ g^{a'b'}( x') \nabla_{a'}\nabla_{b'} -m^2 -\xi
R(x')\right] G \left( x',x \right) =-\delta \left( x',x \right)\; .  \ee $
\delta \left( x',x \right) $ is the delta function on spacetime such that,
for each test function $ f $, \be \label{6} \int d^{4}x' \, \sqrt{-g(x')}
\, f(x') \, \delta \left( x',x \right)  =f(x)  \; .  \ee The retarded
Green function $G_R \left( x',x \right) $ admits the decomposition
\cite{Hadamard}--\cite{Friedlander} 
\be \label{7} 
G_R \left( x',x \right)=
\Sigma \left( x',x \right)\, \delta_R ( \Gamma(x',x) )  + V\left( x',x
\right) \, \Theta_R ( -\Gamma(x',x) ) \; .  
\ee 
$ \Gamma \left(x',x
\right) $ is the square of the proper distance between $ x' $ and $ x $
computed along the unique geodesic connecting $ x' $ and $ x $ in the
normal domain $ \cal{N} $; $\Gamma = 0 $ corresponds to the light cones. $
\delta_R $ and $ \Theta_R $ are, respectively, the Dirac delta
distribution and the Heaviside step function with support in the past of $
x' $. 

The functions $\Sigma$ and $V$ are uniquely determined in a given
spacetime metric \cite{deWittBrehme,Friedlander}. The non--vanishing of
$V( x',x)$ corresponds to the presence of wave tails propagating {\em
inside} the light cone \cite{Hadamard}--\cite{Friedlander}, while the
first contribution to $G_R$, weighted by the coefficient $\Sigma
\left(x',x \right) $ describes sharp propagation {\em along} the light
cone. The structure (\ref{7}) of the retarded Green function is
qualitatively the same for the wave equations satisfied by fields of
higher spin in a curved space. Here, we omit writing these equations and
the corresponding Green functions explicitly, for the sake of brevity. 
However, it is important to remember that the formulation of the Huygens'
principle used in Theorem 1 corresponds to the absence of tails (i.e. $V
\left(x',x \right) =0 $ for all spacetime points $x',x $ in Eq. 
(\ref{7})). Following Ref. \cite{SF93CQG}, one considers a neighborhood $U
\left(x \right)$ of the spacetime point $x \in M$, and one Taylor--expands 
$G_R \left(x',x \right) $, obtaining 
\be \label{sigma}
\Sigma(x',x)=\frac{1}{4\pi}+r_1(x',x)\;, \ee \be \label{vi}
V(x',x)=-\,\frac{1}{8\pi}\left[m^2+\left(\xi-\frac{1}{6}\right)R(x)\right]
+r_2(x',x)\;, 
\ee 
where the remainders $r_{1,2}(x',x) \rightarrow 0 $ as $
x'\rightarrow x $. When the neighborhood $U(x')$ has a small diameter ($
x'\rightarrow x$), there is a tail ($V(x',x) \neq 0 $)  unless the
effective mass $ m_{eff}(x) $ given by \be \label{effectivemass} m_{eff}^2
(x)=m^2+\left( \xi-\frac{1}{6} \right) R(x)  \ee vanishes. We introduce
\\\\ {\bf Definition 1:} the field $\phi $ obeying Eq. (\ref{2})
satisfies the Huygens' principle at the spacetime point $x$ if $ V (x',x)
\rightarrow 0 $ for $ x' \rightarrow x $ in a normal neighborhood of $x
$.\\\\ 
{\bf Definition 2:} the field $\phi $ obeying Eq. (\ref{2})Î
satisfies the Huygens' principle if the latter is satisfied at every
spacetime point $x$.\\\\ Then, a straightforward consequence of Eqs.
(\ref{7}), (\ref{vi}) is\\\\ 
{\bf Lemma:} {\em The solution of Eq.~(\ref{2}) with  $\xi \neq 1/6 $ in
the spacetime $(M,g_{ab})$ satisfies the Huygens'
principle in $x$ if and only if }
\be R(x)=\frac{6m^2}{1-6\xi} \; .  
\ee

The case $\xi=1/6$ is special; in this case there are no tails if and only
if $m=0$, irrespective of the curvature (the value $\xi =1/6 $ is of
physical significance -- see below). We also have\\\\ 
{\bf Theorem 2:} {\em
A sufficient condition for a solution of Eq.~(\ref{2}) with $ \xi \neq 1/6
$ to satisfy the Huygens' principle in the spacetime $(M,g_{ab})$ is
that the latter is a constant curvature space and $ R=6m^2/ (1-6\xi )
$.}\\\\

So far, our considerations have been limited to the mathematical aspects
of the propagation of a scalar field in a curved space. At this point, it
is interesting to examine the subject from the physical point of view. The
physical
reasons for the occurrence of tails are \cite{SF93CQG,BombelliSonego}:\\
{\em i):} The field is massive ($m\neq 0$). For example, the solutions of
the Klein--Gordon equation~(\ref{1}) in the four--dimensional Minkowski
space $(R^4, \eta_{ab})$ have tails whenever $m\neq 0$. \\ 
{\em ii):} The dimensionality of spacetime. For example, the solutions of
Eq.~(\ref{2}) in the $k$--dimensional Minkowski space have tails for odd
$k$, but not for even $k>2$ \cite{Hadamard}. In this paper, we restrict
ourselves to the case of a four--dimensional manifold.\\ 
{\em iii):} Backscattering of the waves off a potential and/or the
spacetime
curvature. This is the most interesting case and in this section we
consider only a non self--interacting field, hence the potential is
absent and we are concerned solely with the backscattering off the
background curvature. The extension to a self--interacting field is
straightforward \cite{SF93CQG}. Moreover, in this paper the dimension of
spacetime is fixed to four, and we are not concerned with tails due to odd
spacetime dimension.

Although the study of the conditions for the absence of wave tails for
massive fields of arbitrary spin (e.g. Refs. \cite{Wunsch,Dowker}) is
legitimate from the mathematical point of view, it is not easy
to justify from the physical perspective. In fact, a field with $m\neq 0$
will have a tail
due the fact that it is massive (this tail is present even in flat space)
and due to the backscattering off the background curvature of spacetime.
The absence of tails means that the two effects exactly cancel each other.
This
situation corresponds to a field with nonzero intrinsic mass that
propagates sharply along the light cone, a phenomenon that has no 
experimental or observational support. A wave tail is indeed a desirable
feature for a massive field; there is not much point in requiring that  
the Huygens' principle be satisfied on a curved space and in deriving the
conditions under hich this ``principle'' is satisfied. As a matter of
fact, these conditions are very restrictive, as is suggested by Theorems 1
and 2. In other words, the Huygens' principle is not a fundamental
principle like, say, the equivalence principle, and its violation is very
realistic.

We conclude this section with an example relevant for cosmology,
which will be used later in Sec. 3. In this example, the balance between
tails due to a mass term and those due to backscattering off the
background curvature is achieved exactly at {\em every} spacetime point.
Keeping in mind Theorem 2, we consider the de Sitter space of constant
curvature $R$, and a test scalar field satisfying Eq.~(\ref{2}), with a
mass
given by 
\be 
m= \left[ R \left( \frac{1}{6} -\xi \right) \right]^{1/2} \ee
for $\xi < 1/6 $. Then $ V (x',x) =0 $ and the field propagates sharply
along 
the light cone at every spacetime point. However, its intrinsic mass $m$
can be made arbitrarily large by suitably choosing the Ricci curvature
(or the constant $\xi $, or both), while the effective mass given by Eq.
(\ref{effectivemass}) vanishes. We
will call this example the {\em ultrapathological} spacetime. Of course,
one could also consider its counterpart obtained by using the anti--de
Sitter space and $\xi > 1/6 $.

\section{Late time mild inflation}

\setcounter{equation}{0}

In this section, we proceed to apply to cosmology the previous
considerations on scalar wave tails. In Ref. \cite{SF93CQG} it was argued
that, if the Einstein equivalence principle \cite{Will} is valid
(i.e. in any metric theory of gravity in which the nature of
$\phi $ is nongravitational), then
in the limit $ x' \rightarrow x $ the solutions of Eq.~(\ref{2}) and the
corresponding Green functions must have the same structure as in flat
space. This corresponds to the local approximation of the spacetime $
(M, g_{ab}) $ with its tangent Minkowski space. The flat space
retarded Green function 
\be 
G_R^{(M)} \left( x',x \right)= \frac{1}{4\pi} \,
\delta_R ( \Gamma(x',x) )  - \left( \frac{m^2}{8\pi}\, + r_3 (x',x) 
\right)  \Theta_R ( -\Gamma(x',x) )  \; , 
\ee 
where $ r_3 (x',x)  \rightarrow 0 $ 
as $ x' \rightarrow x $, must be reproduced in the 
$ x' \rightarrow x $ limit, and this requirement leads to the prescription
$ \xi =1/6 $ for the value of the coupling constant \cite{SF93CQG}. This
result was rederived and confirmed in \cite{GribPoberii,GribRodrigues} and
it can be physically interpreted as the fact that, in the absence of a
scalar field mass, no scale must appear in the local solution to the wave
equation, in analogy with the flat space situation\footnote{Note that
this is not guaranteed by setting $ \xi =0$; in this case the curvature
scale would survive in the Green function, which is the solution for an
impulsive source used in the physical definition of the Huygens' 
principle given by Hadamard \cite{Hadamard}.}. 
The prescription   $\xi =1/6    $ has many consequences for cosmological
inflation. In fact, the
success of many inflationary scenarios strongly depends from the fine
tuning of the
parameter $ \xi $, which is impossible once the value of $ \xi $ is fixed
to the conformal value $ 1/6 $ (\cite{FutamaseMaeda,F96PRD}).

If inflation is driven by a quantum scalar field, the Einstein equivalence
principle probably cannot be imposed. The equivalence principle is likely
to be violated at the quantum level, and the prescription $ \xi=1/6 $ is
not applicable in the quantum regime. However, it is a common belief
that inflation is a classical phenomenon \cite{classinfl}. Moreover, there
are other prescriptions for the value of the coupling constant $ \xi $
(see
references in \cite{F96PRD}) that are valid for quantum fields,
and they differ according to the physical nature of the field
$ \phi $. The existence of tails of radiation, and the issue of the value
of $ \xi $ are relevant also for other areas of cosmology and of
theoretical physics
\cite{EllisSciama,FS92PLA,HochbergKephart,Hiscock,F96PRD}.

Currently, cosmology faces two problems raised by recent observations: the
age of the universe problem (the age of certain globular clusters is
larger than the age of the universe inferred from the method of Cepheid
variables \cite{Pierce,Freedman,Mould}),
and the discrepancy between the
local and the global (based on the Zeldovich--Sunyaev effect
\cite{Jones,Birkinshaw}) measures of the Hubble parameter $ H_0 $. In
order to reconcile theory and observations, it has been proposed
that the universe undergoes short periods of piece--wise exponential
expansion
that interrupt the matter--dominated era after star formation (``late time
mild inflation'' or LTMI) \cite{LTMI}. 

\subsection{The proposed mechanism for LTMI is physically pathological}

The mechanism proposed in \cite{LTMI} to achieve LTMI is based on a
classical, massive, non self--interacting scalar field
nonminimally coupled to the Ricci curvature of spacetime, and satisfying
Eq.~(\ref{2}), in the context of general relativity. The authors of Ref.
\cite{LTMI} assume a Einstein--de Sitter universe and a baryon density of
order $ \Omega_m=0.01 $ (in units of the critical density $
\rho_c=3H^2/8\pi G $). The Einstein equations for a mixture of dust and a
scalar field are 
\be \label{H} 
H^2=\frac{8\pi G}{3} \left( \rho_m +
\rho_{\phi} \right) \; , 
\ee 
\be \label{HH} 
\dot{H}+H^2 = - \frac{4\pi G}{3} \left( \rho_m + \rho_{\phi} +3 P_{\phi}
\right) \; , 
\ee 
where $ H =
\dot{a}/a $ is the Hubble parameter, $a(t)$ is the scale factor of the
Einstein--de Sitter line element, and a overdot denotes differentiation
with respect to the comoving time $t$. $ \rho_{m} $ is the energy density
of dust, $P_m=0 $, and the energy density and pressure of the scalar field
component of the cosmic fluid are given by \be \label{density}
\rho_{\phi}=\left( 1-8\pi G \xi \phi^2 \right)^{-1} \left[ \frac{(
\dot{\phi})^2}{2}+F( \phi)+6\xi H \phi \dot{\phi}\right] \; , \ee \be
\label{pressure} P_{\phi}=\left( 1-8\pi G \xi \phi^2 \right)^{-1} \left[
\left( \frac{1}{2}-2\xi \right) \dot{\phi}^2-F( \phi)-2\xi \phi
\ddot{\phi}-4\xi H \phi \dot{\phi} \right] \;, \ee respectively, where $
F( \phi )=m^2 \phi^2 /2 $. The Klein--Gordon equation becomes 
\be
\ddot{\phi} +3H \dot{\phi}+m^2 \phi +6\xi \left( \dot{H}+2 H^2 \right) 
\phi=0 \; .  
\ee 
A period of LTMI corresponds to the particular solution
\be H_*=\left( \frac{m^2}{12|\xi |} \right)^{1/2} \; , \;\;\;\;\;\;\;\;
\phi_*^2 = \frac{1}{8\pi G |\xi |} \; ,  
\ee 
for which $ \rho_{\phi} \geq 0 $, $ P_{\phi}=-\rho_{\phi} $. Due to the
onset of instabilities, the exponential expansion soon stops and is
followed by a oscillatory decay (due to the fact that $ m \neq
0 $).
The values of the parameters $ m $ and $ \xi $ have to be adjusted in
order to fit the observations; in particular, a negative value of $ \xi $
and a rather large (compared to unity) value of its modulus are essential
for a successful LTMI \cite{LTMI}. The authors of Ref. \cite{LTMI} chose $ \xi
=-80 $ and $ m=10^{-31} $ eV (although this is more an example than a best
fit of the observational data, it gives an idea of the orders of
magnitude of the parameters $\xi, m$ needed for an interesting LTMI).

In the light of the result of Ref. \cite{SF93CQG} explained at the
beginning of this section, the value of the coupling constant $ \xi $ is
fixed to
$1/6$ in general relativity, and the LTMI scenario does not work. It is in
principle possible that LTMI can be achieved in the context of a theory of
gravity and of the boson field in which the prescription $ \xi = 1/6 $ does
not apply. In this case, one still has to deal with the other
prescriptions for
the value of $ \xi $ existing in the literature (see \cite{F96PRD} for a
review). A more serious problem is that, even ignoring the prescription $
\xi =1/6 $ coming from the Einstein equivalence principle, each phase of
LTMI is extremely close to the ultrapathological spacetime described at
the end of 
the previous section. In fact, LTMI corresponds to the vanishing of the
effective mass given by 
\be \label{effectivemass2} 
\mu^{2}= m^{2} +6 \xi \left(\dot{H} +2 H^2 \right) \; , 
\ee 
while the
ultrapathological space
corresponds to the vanishing of $ m_{eff} $ given by Eq.
(\ref{effectivemass}), 
\be 
m_{eff}^{2}= m^{2} +\left( 6\xi -1 \right) 
\left( \dot{H} +2 H^2 \right)  \; .  
\ee 
For $ \xi >> 1 $, $ m_{eff}
\simeq \mu $ and LTMI essentially reproduces the ultrapathological space.
Using the value $ \xi =-80 $ of Ref. \cite{LTMI}, one obtains from Eq.
(\ref{effectivemass2}) that $ H_*^{2} \simeq 1.0416 \cdot 10^{-3} m^{2} $
, while the ultrapathological case corresponds to $ H_*^{2} \simeq 1.0395
\cdot 10^{-3} m^{2} $. A very substantial part of the tail of $ \phi $ due
to the intrinsic mass $m$ is cancelled by the tail due to the
backscattering off the background curvature of spacetime. The cancellation
becomes more and more precise as $ - \xi $ increases, which makes
inflation more and more pronounced \cite{LTMI}.

\subsection{Alternative mechanisms for LTMI}

The mechanism used in Ref. \cite{LTMI} to achieve LTMI is clearly
unphysical. Is there a realistic mechanism that works ? In order to answer
this question, one possibility is adding
a nontrivial (i.e. not a pure mass term) potential $ F( \phi ) $ to the
picture. However, one then defines the intrinsic mass of the scalar field
in the late time mild inflationary state $(H_*, \phi_* )$ as $ m_{\phi} =
d^2 F/d \phi^2 ( \phi_* ) $, and one is facing again the problem of
the cancellation between the tail due to the intrinsic mass $m$ and the
tail due to the backscattering off the background curvature.

A way out of
this dilemma could be the consideration of the linear potential $F( \phi
)=\lambda \phi$, for which $ m_{\phi} = 0 $. In this case, the Einstein
equations (\ref{H}), (\ref{HH}), supplemented by the expressions for the
energy density and pressure of the scalar field (\ref{density}),
(\ref{pressure}) admit, for $  \xi    <0 $, the de Sitter solution 
\be
H_{**}^2=\left( \frac{\pi G}{6| \xi |}\right)^{1/2} \lambda \; ,
\;\;\;\;\;\;\;\; \phi_{**}^2 = \frac{1}{24\pi G | \xi |} \; .  
\ee
In principle, one can stop a late time inflation of this kind; in the
original mechanism for LTMI proposed in Ref. 
\cite{LTMI}, the exit from the exponential expansion was due to the
Ljapunov instability of the de Sitter solution against small
perturbations. For a field in a linear potential the de Sitter solution is
also unstable. In fact,
consider the universe in a state that is a small perturbation of the
$(H_{**}, \phi_{**} )$ inflationary state, 
\be 
\phi = \phi_{**} ( 1+x ) \; , \;\;\;\;\;\;\;\;  H=H_{**} ( 1+y )  \; ,  
\ee
where $ x$ and $y$ are small
compared to unity. After straightforward calculations, one obtains the
evolution equations for the perturbations, 
\be \left( \begin{array}{c}
\dot{x} \\ \dot{y} \end{array} \right) =\left( \begin{array}{cc} a_1 & a_2
\\ a_3 & a_4 \end{array} \right) \left( \begin{array}{c} x \\ y
\end{array} \right) \; , \ee where 
\be 
a_1= \alpha \; , 
\ee 
\be
a_2=a_4=-4\alpha \; , \ee \be a_3=-\, \frac{2\alpha}{3\xi -2} \; , 
\ee 
\be
\alpha =\left( \frac{\pi G}{6 |\xi |} \right)^{1/4} \lambda^{1/2} \; ,
\ee or 
\be 
\dot{\underline{x}} = \alpha M \, \underline{x} \; .  
\ee 
The
matrix $M$ has real eigenvalues 
\be 
s_{1,2} = \frac{3}{2} \left( -1 \pm
\sqrt{1-\frac{16 \xi}{2-3\xi}} \right)  \; , 
\ee 
where the discriminant $ \Delta = (18-75\xi )(2-3\xi)^{-1} > 0 $ for $
\xi < 0 $. Since $ s_{1} $, $ s_{2} $ have opposite signs, $ \left(
H_{**}, \phi_{**}
\right) $ is a saddle point, and describes an unstable equilibrium. A
perturbation  can grow and break the exponential expansion.

Another possibility that one can naturally think of in order to
avoid the ultrapathological spacetime, consists in requiring that, 
during LTMI, the growth of the scale factor be accelerated, but not
exponential. For
example, one can search for piece--wise power--law inflationary solutions
$a=a_0 t^p $, where $p>1 $. Then, the Ricci curvature is not constant and
the exact cancellation of mass and curvature tails can occur at most at a
single instant in the history of the universe; because of the monotonic
behaviour of the Ricci curvature $ R=6p( 2p-1) t^{-2} $, the equation $
m_{eff}=0 $ has only one root. This instant of pathological behaviour can
be avoided by making the piece--wise period of inflation sufficiently
short.  A power--law inflationary solution for a universe driven by a
nonminimally coupled scalar field with the potential 
\be     \label{polpotential} 
F( \phi )= A \phi^n \; , \;\;\;\;\;\;\;\;\; n>6
\; , 
\ee 
was found in Ref. \cite{FutamaseMaeda}. One has, for this
solution, 
\be 
p= 2\, \frac{1+(n-10) \xi}{(n-4)(n-6)
 | \xi | } \; .  
\ee 
In general relativity, the prescription $ \xi=1/ 6 $ yields $ p= 2/ (n-6)
$, which corresponds to {\em i)} accelerated expansion if $ 6< n <8 $,
{\em ii)} to a coasting
universe if $ n= 8 $, and {\em iii)} to a decelerated universe which still
expands faster than $ a(t)=a_0 t^{2/3} $ if $ n < 9  $. Hence, in
principle, one can
achieve periods of LTMI in general relativity, with the potential
(\ref{polpotential}). The detailed analysis of these alternative
mechanisms for LTMI and their
comparison with the cosmological observations are beyond the scope of
this paper, which focuses on tails of radiation. In addition,
it would be desirable to identify the
scalar field in the potential (\ref{polpotential}) with some known field
from high energy physics, which is not done in the phenomenological
approach to LTMI.

\section{Discussion and conclusions}

\setcounter{equation}{0}

The violation of the Huygens' principle and the presence of tails of
radiation have been studied for many years in the context of mathematical
physics. Only recently it has been realized that tails of radiation have
important physical applications. An example in astrophysics is given by
the
tails in the gravitational radiation emitted by compact objects. These
tails are relevant for the correct data analysis (matched filtering) in
the large laser interferometric detectors of gravitational waves ({\em
LIGO, VIRGO, GEO600, TAMA, ...}) \cite{gwtails}. 

In classical field theory in curved spaces, a counterintuitive result is
that
the absence of pathologies in the propagation of scalar fields fixes to $
1/6 $ the value of the coupling constant $ \xi $ of the scalar field with
the
Ricci curvature \cite{SF93CQG}.  This prescription has far--reaching
consequences for cosmological inflation (\cite{F96PRD} and references
therein), for the cosmic no--hair theorems \cite{Starobinski,FRM}, and
possibly for other areas of cosmology and of theoretical physics
\cite{F97GRG,F96PRD,Hiscock}.

In the present paper, we have considered
the idea of LTMI, recently proposed to solve the age of the
universe problem and the puzzle of the discrepancy between the local and
global measures of the Hubble parameter. While the idea of
LTMI appears to be very valuable, unfortunately the mechanism employed to
achieve it (a
classical, massive, non self--interacting scalar field nonminimally
coupled to the Ricci curvature) is not viable, because it corresponds
to
the extremely pathological physics discussed in Sec. 2. In
fact, the LTMI scenario almost exactly reproduces the
ultrapathological spacetime. Alternative mechanisms to generate LTMI 
are discussed in Sec. 3, and the possibility of
having LTMI with a negatively coupled, self--interacting scalar field is
not ruled out in general relativity. However, one must be willing
to pay the price of
introducing a suitable scalar field potential and obtaining a
less--than--exponential expansion of the universe during LTMI. 

The approach to LTMI is purely phenomenological, and no
serious attempt is made to identify the scalar field with a known field
from a high energy physics theory. The hypotetical possibility of
identifying the scalar field with a superlight Proca field \cite{LTMI} 
clearly does not work if the field is self--interacting (apart from the
problem that a homogeneous vector field would introduce anisotropy, and a
nontrivial distribution of this vector field would need to be considered
\cite{LTMI}). The analysis of the LTMI scenario would have to be
redone if a vector field instead of a scalar one was used as a source term
in the right hand side of the Einstein equations. On the other hand,
fields of different spin in the same background metric have different
behaviour with respect to tails. For example, the Maxwell field satisfies
the Huygens' principle in a Friedmann--Robertson--Walker space; in fact,
the latter is conformally flat, and the Maxwell equations are conformally
invariant. The tail--free property of the Maxwell field in Minkowski space
is then transferred to the Friedmann--Robertson--Walker space
\cite{SF92JMP,Noonan}. Hopefully,  a viable mechanism will be found which
is capable of successfully implementing the idea of LTMI. Work in this
direction is in progress.

The model of the universe analogous to that of LTMI, but with $\xi
> 0 $, does not give rise to inflation and was considered in Ref.
\cite{Morikawa} in order to explain the reported periodicity in the
redshift of galaxies \cite{BEKS}.  A look at the latter model with the
knowledge of scalar field tails is also instructive \cite{F97GRG}, and
leads to information on the nature of the correct theory of gravity,
should the reported redshift periodicity turn out to be genuine and not an
artifact of incomplete or faulty statistics.

Previous literature \cite{F97GRG} and the present paper show that tails of
scalar fields and nonminimal coupling to the Ricci curvature are very
relevant for cosmology, and not only for inflationary theories. Moreover,
tails and nonminimal ($\xi \neq 0 $) coupling are forced upon us in almost
all situations of physical interest. Thus, it is seen that the study of
these phenomena is not optional, rather it is necessary in cosmology.

\section*{Acknowledgments}

V.F. is grateful to Varun Sahni for stimulating discussions, to the
{\em RggR} group for a visit at the Universit\'{e} Libre de Bruxelles,
where this paper was begun, and to L. Niwa for a reading of the
manuscript. This work was partially supported by EEC
grants numbers PSS*~0992 and CT1*--CT94--0004, and by OLAM, Fondation pour
la Recherche Fondamentale, Brussels. 

\end{document}